\documentclass[aps,prl,twocolumn,nofootinbib,notitlepage,superscriptaddress,aps,pra]{revtex4-1}
\usepackage[utf8x]{inputenc}
\usepackage{float}
\usepackage{amsmath}
\usepackage{amssymb}
\usepackage{graphicx}
\usepackage{color}



\begin{document}
\title{Phenomenology of curvature-induced quantum-gravity effects}
\author{Giovanni AMELINO-CAMELIA}
\affiliation{Dipartimento di Fisica Ettore Pancini, Università di Napoli ”Federico II”, and INFN, Sezione di Napoli, Complesso Univ. Monte S. Angelo, I-80126 Napoli, Italy}
\author{Giacomo ROSATI}
\affiliation{Institute for Theoretical Physics, University of Wroclaw, Pl.\ Maksa
Borna 9, Pl--50-204 Wroclaw, Poland}
\author{Suzana BEDI\'C}
\affiliation{ICRANet
, P.le della Repubblica 10, 65100 Pescara, Italy,\\ and
ICRA and University of Rome ``Sapienza'', Physics Department, P.le
A. Moro 5, 00185 Rome, Italy}
\begin{abstract}
\noindent Several studies have been devoted to the possibility that quantum gravity
might tangibly affect relativistic kinematics for particles propagating
from distant astrophysical sources to our telescopes, but the relevant literature
has so far focused exclusively on a subclass of scenarios such that the
quantum-gravity effects are independent of (macroscopic) curvature. It was assumed
that a phenomenology for quantum-gravity
effects that are triggered by curvature might be a dead end
because of a double suppression: by the smallness of the characteristic
quantum-gravity length scale and by
the smallness of curvature. This state of affairs
is becoming increasingly unsatisfactory in light of some recent quantum-gravity studies
providing evidence of the fact that the presence of curvature might be required in order
to have the novel relativistic properties. We here analyze an explicit
scenario for curvature-induced quantum-gravity effects, and show that
the smallness of curvature does not pose a challenge for phenomenology since it is compensated
by the large distances traveled by the particles considered in the relevant phenomenological studies.
We also observe that the present data situation for particles
propagating from distant astrophysical sources to our telescopes,
while inconclusive, provides more encouragement for our curvature-induced
effects than for the curvature-independent effects that were so far
studied.
\end{abstract}
\maketitle

We will soon reach a full century of theoretical research
on the quantum-gravity problem, since the first such studies date
back to the early 1930s~\cite{bronstein,BornReciprocity}, but the development of a quantum-gravity phenomenology is only rather recent
 and still focuses on very few opportunities (see Ref.~\cite{GACphen} and references therein). This timeline mismatch is due to the horrifying smallness of the Planck length, the scale expected to be characteristic of quantum-gravity effects, which only in very rare
observational situations could lead to a tangible imprint in data. The most studied  opportunity of this sort concerns
the possibility that the
laws of relativistic kinematics might be
 deformed by quantum gravity, in ways that could leave
a trace in the analysis of gamma-ray bursts (GRBs). Since GRBs are sourced at very large distances (often
at redshift between 0.5 and 4) the minute quantum-gravity
effects might accumulate to observably-large magnitude along the way~\cite{GACphen}.

In light of the rare opportunity that these GRB studies provide, a large literature has been devoted to them (see, e.g., Ref.~\cite{PiranRodriguez,JacobPiran,IceCubeNature,AmelinoCamelia:1997gz,schaefer,scargle,Abdo:2009zza,JacobPiran,Ackermann2009,Bolmont:2010np,Vasileiou,Zhang2014,xuma2016_2}), but all these studies
have  focused  on a subclass of scenarios such that the
quantum-gravity effects persist even in absence of (macroscopic) curvature.
This affords some technical simplifications, and it was so far assumed that
anyway a phenomenology of quantum-gravity
effects that are triggered by curvature would be hopeless, since it would
have
to face not only the smallness of the characteristic length scale of quantum-gravity but also
the smallness of  curvature.
However, on the theory side one finds quite some motivation for considering
quantum-gravity
effects that are triggered by curvature: for example, models of quantum spacetime based on Hopf algebras
typically produce a mechanism such that the presence of quantum-gravity effects requires curvature,
with the possibility of curvature-independent quantum-gravity effects arising only at the cost
of some sizeable fine tuning~\cite{GACsmolinStaro,aschieri09},
and some lines of analysis based on the loop-quantum-gravity
perspective suggest that the relevant quantum-gravity effects must be triggered by curvature~\cite{bianchiRovelli}.

We here report a study in which we chose to focus on the possibility
of quantum-gravity
effects that are triggered by curvature. We got started wanting to quantify the ``cost''
of the curvature suppression but remarkably we found that there is no suppression at all: essentially
curvature appears in the relevant formulas always multiplied by the distance traveled by the particle,
which indeed in the relevant phenomenology is huge.

For this first exploratory study we were satisfied by considering a specific
model with effects triggered by curvature. We do not exclude that our model might deserve
intrinsic interest, but for the purpose of our analysis it should be viewed as a toy model used
to establish that a phenomenology of curvature-induced quantum-gravity effects is possible.
Our starting point is the model which has been most studied in the relevant phenomenology,
whose core feature is the following dispersion relation (see, e.g., Refs.~\cite{JacobPiran,DSRFRW}
and references therein)
\begin{equation}
E=\frac{p}{a\left(t\right)} \left( 1 -\frac{\lambda}{2}\frac{p}{a\left(t\right)} \right).
\label{eq1}
\end{equation}
where $E$ is the ``comoving-time energy'' ($E \sim \partial_{t}$~\cite{DSRFRW}),
$a(t)$ is the scale factor in FRLW spacetime, $p$ is
the modulus of the (comoving, constant) momentum measured by an observer at
the detector (which coincides with the redshifted photon energy measured at the detector), and $\lambda$ is a phenomenological length scale characterizing
the quantum-spacetime effects, to be determined experimentally.

An alternative which has also attracted some interest~\cite{PiranRodriguez,DSRFRW}
is
\begin{equation}
E=\frac{p}{a\left(t\right)} \left( 1 -\frac{\lambda'}{2}p \, a\left(t\right) \right).
\label{eq2}
\end{equation}
Both (\ref{eq1}) and (\ref{eq2}) are such that in-vacuo dispersion would be present
even without curvature; however, as we shall soon show, their combination
\begin{equation}
E=\frac{p}{a\left(t\right)} \left( 1 -\frac{\lambda}{2}\frac{p}{a\left(t\right)} -\frac{\lambda'}{2}p \, a\left(t\right) \right).
\label{eq3}
\end{equation}
produces no dispersion in absence of curvature, if $\lambda'=-\lambda$. For the purposes of this study
one could view (\ref{eq3}) simply as a toy model
well suited for a first exploration of the possible size of curvature-triggered effects,
a model whose simplicity and connection to scenarios studied previously
might provide a good stepping stone for future more refined studies of these issues.
In the following we shall derive results that depend on both $\lambda'$ and $\lambda$,
but our main interest will be in the case $\lambda'=-\lambda$.

From (\ref{eq3}) it follows that the (comoving-time) velocity
is
\begin{equation}
v\left(t\right)=\frac{1}{a\left(t\right)} \left(1- \lambda \frac{p}{a\left(t\right)} -\lambda' p \, a\left(t\right)\right).\label{velocity}
\end{equation}
This can be usefully rewritten as a relationship between  velocity and redshift $z=1/a\left(t\right)-1$,
working with a scale factor such that $a\left(0\right)=1$, {\it i.e.}
the scale factor
is 1 at the time of detection. One easily finds that
\begin{equation}
v\left(z\right)=1+z-\left(\lambda+\lambda'+2\lambda z\left(1+\frac{1}{2}z\right)\right)p.\label{velocityRedshift}
\end{equation}
The fact that for $\lambda'=-\lambda$ the  correction is proportional
to the redshift signals   that indeed for $\lambda'=-\lambda$ the dispersion effects are
a purely ``curvature-induced'' correction. The feared possibility that the smallness of curvature might lead to
unobservably small effects finds confirmation only for observations conducted on small distances: for small
redshift the curvature-induced effects on velocity are much smaller than the ones found when $\lambda' \neq -\lambda$;
however, the relevant phenomenology focuses on GRBs at redshifts of order unity or bigger, and in that case the magnitude
of our curvature-induced effects is comparable to that of the effects produced when $\lambda' \neq -\lambda$.
So we proved that
in the phenomenology of GRBs the feared suppression of curvature-induced effects is not present.

Further insight can be gained by rewriting the formula for the velocity using
 the relation between the redshift and the Hubble constant
$H_{0}=\dot{a}/a\Big|_{t=0}$, derived from Taylor expanding the
redshift equation for small distances (i.e. small times $t$), $z\simeq1/(1+\dot{a}t)-1\simeq-H_{0}t$. This gives
\begin{equation}
v\left(t\right)\simeq 1 - H_0 t -\left(\left(\lambda+\lambda'\right)-2\lambda H_{0}t\right)p.
\end{equation}
Here again we see that the dispersion effects disappear if $\lambda'=-\lambda$ and there is no curvature ($H_{0} =0$);
moreover, the curvature-induced effects are only significant for large distances (large values of $t$).

The GRB analyses used for these studies of quantum-gravity-induced dispersion of course do not
measure directly the velocity but rather the difference in time of arrival for particles of different energies
which one can presume to have been emitted nearly simultaneously~\cite{GACphen}.
We therefore need to compute the amount
of time-of-arrival modification produced by our formulas for velocity modification. This is easily done
 by integrating equation (\ref{velocityRedshift})
from the redshift of the source $z_{em}$ to $z_{now}=0$, considering
that $dt=-dz/\left(H\left(z\right)\left(1+z\right)\right)$, and one finds
\begin{equation}
\label{deltat}
\Delta t=\Delta p \int_{0}^{z_{em}}\frac{dz}{H\left(z\right)}\left(\lambda\left(1+z\right)+\frac{\lambda'}{\left(1+z\right)}\right),
\end{equation}
which of course for $\lambda'=-\lambda$ describes effects that are very small at small redshift:
\begin{equation}
\label{deltatCurv}
\Delta t \big|_{\lambda'=-\lambda}=2\lambda\Delta p \int_{0}^{z_{em}}\frac{dz}{H\left(z\right)}\frac{z+z^{2}/2}{1+z} \, .
\end{equation}
For the Hubble parameter we adopt the favored description of current cosmological models, {\it i.e.}  $H\left(z\right)=H_{0}\sqrt{\Omega_{m}\left(1+z\right)^{3}+\Omega_{\Lambda}}$
where $H_{0}$, $\Omega_{m}$ and $\Omega_{\Lambda}$ are respectively the Hubble constant, the matter and the cosmological constant density parameters (to which we assign values according to Ref.~\cite{planckParam}),
and $\Delta p$ is the difference between the comoving (constant) momentum of two (simultaneously-emitted)
particles whose times of arrival are being compared.

In order to render more tangible our findings concerning the viability of the
phenomenology of curvature-triggered quantum-spacetime effects, we find appropriate to consider some data whose
analysis in terms of the  $\lambda'=0$ scenario (the standard scenario with effects that do not turn off in the zero-curvature limit)
has been attracting some attention. Specifically, we take as reference the findings for the  $\lambda'=0$ scenario
reported in Refs.\cite{xuma2016_2,IceCubeNature}, whose selection criteria led the authors to focus on 11 GRB photons of particularly high energy
observed by the Fermi telescope.
The key properties of those 11 photons are shown in table~\ref{tableGRB}.
A detailed description of the selection criteria can be found in Refs.\cite{xuma2016_2,IceCubeNature}, with the most noteworthy among
them being those concerning the requirements on energy of the photon and the time window allowed for
considering a photon as possibly being emitted in coincidence with the first peak:
the selected photons should be such that their energy at emission (factoring in redshift) is greater than 40 GeV, and
 should be compatible with an intrinsic time lag (``inferred offset at the source")
 with respect to the first low-energy peak of the GRB of no more than 20 seconds.

\begin{table}[htbp]
\centering
{\def\arraystretch{0.5}\tabcolsep=3pt
\begin{tabular}{|lccl|}
\hline
   ~ event ~~~~ & $z$ & $E_{\text{obs}}$ [GeV] & $ \Delta t $ [s]  \\ \hline
130427A & 0.34  & 77.1 &18.10 \\
090510  & 0.90  & 29.9 & 0.86  \\
160509A & 1.17  & 51.9 & 62.59 \\
100414A & 1.37  & 29.8 & 33.08\\
090902Ba & 1.82  & 14.2 &  4.40\\
090902Bb & 1.82  & 15.4 & 35.84\\
090902Bc & 1.82  & 18.1 & 16.40\\
090902Bd & 1.82  & 39.9 & 71.98\\
090926A & 2.11  & 19.5 & 20.51\\
080916Ca & 4.35  & 12.4 & 10.56\\
080916Cb & 4.35  & 27.4 & 34.53\\ \hline
\end{tabular}
}
\caption{\label{tableGRB}
Data taken from the Fermi gamma-ray space observatory,
 publicly available at
https://fermi.gsfc.nasa.gov/ssc/data/access/, concerning
 the 11 photons featured in our figures. The first column reports the name of the events, specified by the GRB name followed by a (lowercase) letter in case of multiple photons associated to the same GRB.
The second column reports the redshift of the relevant GRB. The third column reports the (observed) value of events energy at the detector. The last column reports the difference in times of observation between the photon and first low-energy peak of the GRB.}
\label{table1}
\end{table}

Notably, the objectives of those analyses of Refs.~\cite{xuma2016_2,IceCubeNature}
 were in part achieved: after fitting on data
just two parameters ($t_{off}$,
accounting for an average time offset at the source~\cite{IceCubeNature,xuma2016_2,shaoLag}, and $\lambda$),
one finds that 8 out of the 11 data points show a delay of arrival, with respect to the first peak of the relevant GRB,
consistent with the  $\lambda'=0$ scenario.
The fact that 3 out of the 11 data points do not fit the model
is not necessarily concerning, since the assumption that the highest-energy photons be emitted in coincidence
with the first bright low-energy peak can at best work most of the time, but surely not always.
In particular, the fact that the data point 090902Bb reflects a larger delay than predicted by the model might just
mean that it was emitted in coincidence with a later peak, not the first peak.
The data point 080916Cb reflects a smaller delay than predicted by the model (and emission by peaks later than the first one can
only inflate the delay estimate, not reduce it); however, the data point 080916Cb is affected by rather large uncertainties.
The main, and significant, challenge to the interpretation
of Fig.1 in connection with the $\lambda'=0$ scenario resides in the data point 090510, whose delay with respect to
the time of observation of the first low-energy peak is much smaller than predicted by the $\lambda'=0$ scenario,
and it is a data point with very little uncertainty. The data point 090510 cannot be described in any way
within the $\lambda'=0$ scenario.

\begin{figure}[h!]
\centering{}\includegraphics[scale=0.35]{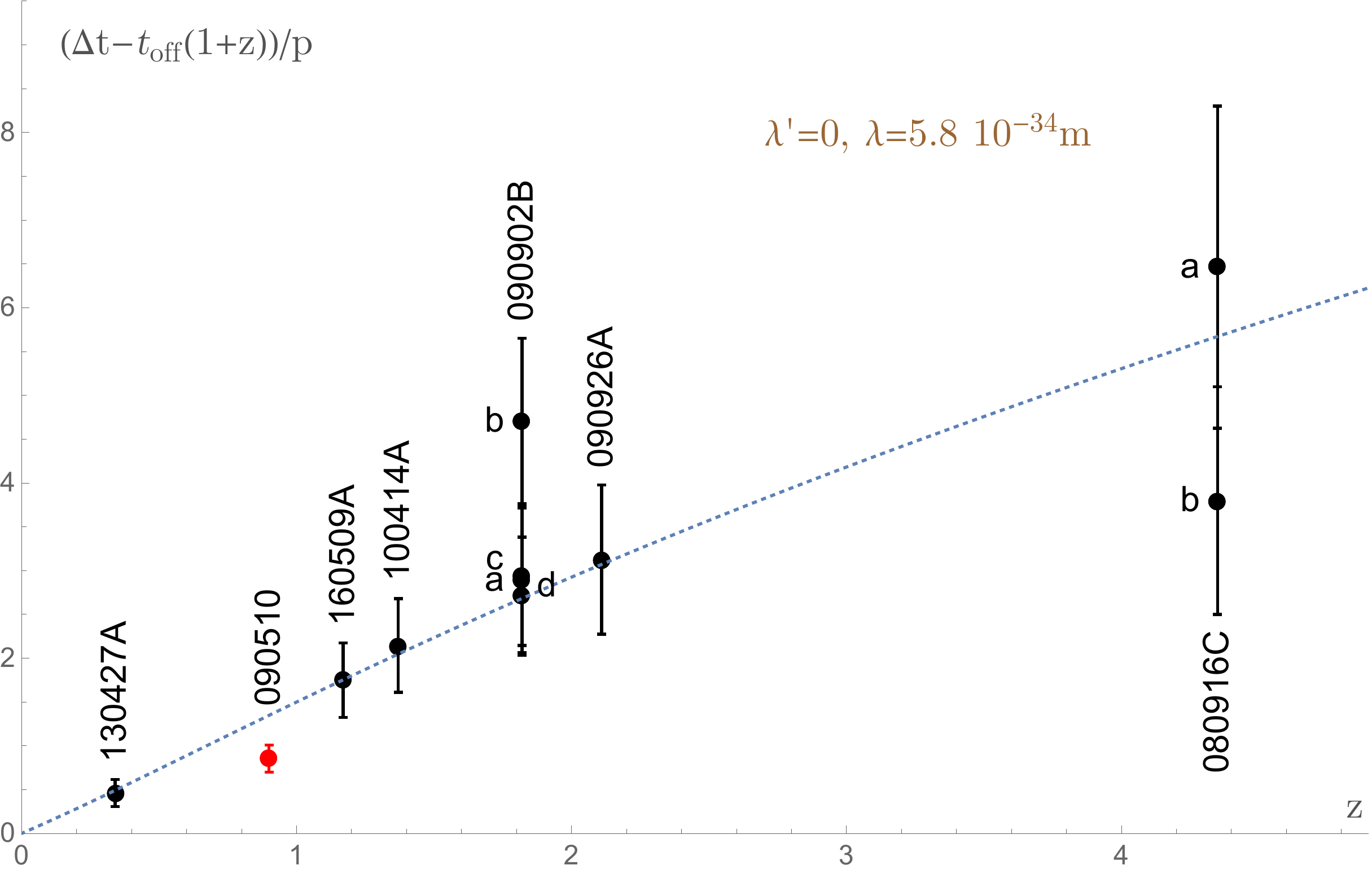}\caption{
\label{fig1}
The  11 photons already considered in Refs.~\cite{xuma2016_2,IceCubeNature},
here characterized in terms of the values of their $\left(\Delta t-t_{off}(1+z)\right)/ p$ and their redshift $z$, two variables which,
according to Eq.~(\ref{deltat}), should be linked by the dotted blue curve, if $\lambda'=0$ and $\lambda = 5.8 \, 10^{-34} m$.}
\end{figure}

Fig.2 reports the results for the same analysis redone for our
$\lambda'=-\lambda$ scenario with
curvature-triggered effects. First of all one should notice that, in agreement with our main thesis, for comparable values
of $\lambda$ our novel $\lambda'=-\lambda$ scenario and the much studied $\lambda'=0$ scenario both produce effects
at the level needed for this sort of analysis: there is no overall suppression of the effects due to the smallness
of spacetime curvature. Intriguingly, the $\lambda'=-\lambda$ scenario predicts much weaker
 in-vacuo-dispersion effects only for small redshifts, with an impact on the data analysis which however
 is encouraging: as seen by comparing Fig.2 to Fig.1, the $\lambda'=-\lambda$ scenario provides
 a description of the data which overall is as good as that of the $\lambda'=0$ scenario, and the $\lambda'=-\lambda$ scenario
 handles  nicely also the 090510 data point, the one which is instead very problematic for
 the $\lambda'=0$ scenario.

\begin{figure}[H]
\centering{}\includegraphics[scale=0.35]{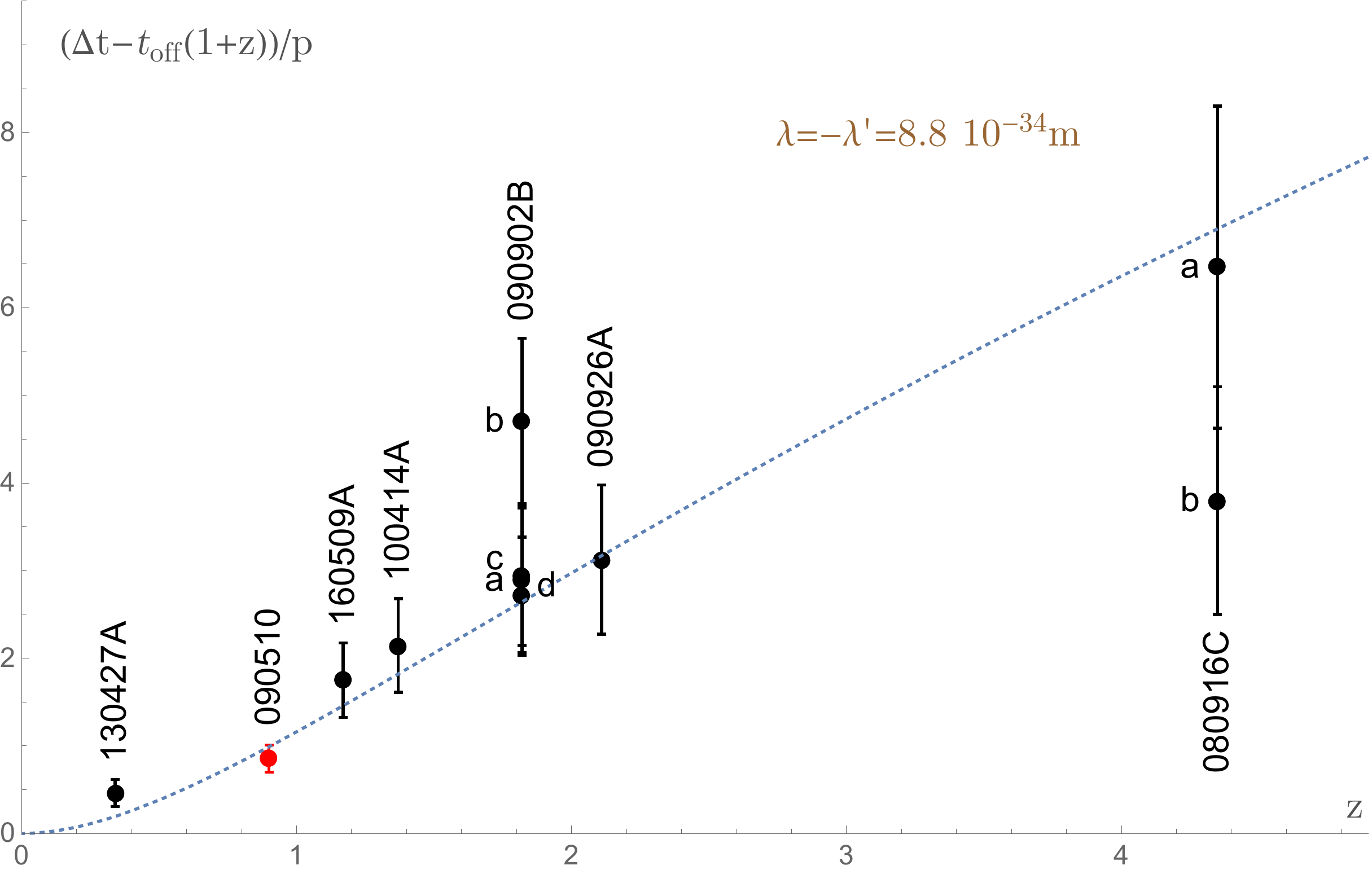}\caption{
\label{fig2}
Same as fig.\ref{fig1}, but here the dotted blue curve is for
 Eq.~(\ref{deltat}) in the ``curvature-induced case'' $\lambda'=-\lambda $, with
 $\lambda = 8.8 \, 10^{-34} m$.}

\end{figure}

We shall not dwell further on these data. They are intriguing, particularly when analyzed in relation to
our novel $\lambda'=-\lambda$ scenario, but it is after all a small data set, with several data points affected
by large uncertainties. More events relevant for such analyses will be observed by our telesopes,
and chances are the relevant feature will fade away as more data points are gathered.
We have however established robustly that a phenomenology of curvature-triggered quantum-gravity effects
is viable. Our  $\lambda'=-\lambda$ scenario might well turn out to deserve intrinsic interest, but
we introduced it here only as a tool for establishing the viability of this new phenomenology.
Some of the features uncovered with our $\lambda'=-\lambda$ scenario (particularly the slow onset of effects at
small redshifts) are likely to be shared by other scenarios for curvature-triggered quantum-gravity effects,
but of course there will be model-dependent aspects of the phenomenology.
The success of this
phenomenology will therefore depend not only on Nature's indulgence, but also on our ability to identify efficacious
models of curvature-triggered quantum-gravity effects, providing guidance to the phenomenology.
We feel this should motivate the community, particularly those with expertise in spacetime noncommutativity and loop quantum gravity,
to adopt a more quantitative (more phenomenological) approach toward the analysis of
the structure of curvature-triggered effects emerging in different quantum-gravity theories.

\section*{Acknowledgements}

G.A.-C.'s work on this project was supported by the FQXi grant 2018-190483 and by the MIUR, PRIN 2017 grant 20179ZF5KS.
G.R.'s work on this project was supported by the Polish National Science Centre grant 2019/33/B/ST2/00050.
This work also falls within the scopes of the EU COST action CA18108 ``Quantum gravity phenomenology in the multi-messenger era".
We gratefully acknowledge valuable conversations with Paolo Aschieri, Luciano Burderi, Irene De Vico Fallani, Fabrizio Fiore,
Giancarlo Ghirlanda, Tsvi Piran and Andrea Sanna.


\begin{thebibliography}{50}

\bibitem{bronstein}
M.~Bronstein
Phys. Z. Sowjetunion {\bf 9} (1936) 140;
Gen.\ Rel.\ Grav.\  {\bf 44} (2012) 267.

\bibitem{BornReciprocity}
M.~Born,
Proc. R. Soc. A \textbf{165} (1938) 291.

\bibitem{GACphen}
  G.~Amelino-Camelia,
  Living Rev.\ Rel.\  {\bf 16} (2013) 5.

\bibitem{AmelinoCamelia:1997gz}
G.~Amelino-Camelia, J.~R.~Ellis, N.~E.~Mavromatos, D.~V.~Nanopoulos and S.~Sarkar,
Nature \textbf{393} (1998), 763-765.

\bibitem{schaefer}
B.~E.~Schaefer,
Phys. Rev. Lett. \textbf{82}, (1999), 4964-4966.

\bibitem{PiranRodriguez}
M.~Rodriguez Martinez and T.~Piran,
JCAP \textbf{04} (2006) 006.

\bibitem{scargle}
J.~D.~Scargle, J.~P.~Norris, and J.~T.~Bonnell,
Astrophys. J. \textbf{673}   (2008) 972-980.

\bibitem{JacobPiran}
U.~Jacob and T.~Piran,
JCAP \textbf{0801} (2008) 031.

\bibitem{Abdo:2009zza}
  A.~A.~Abdo {\it et al.}  [Fermi GBM/LAT],
  Science {\bf 323} (2009) 1688.

\bibitem{Ackermann2009}
A.~A.~Abdo \textit{et al.} [Fermi GBM/LAT],
Nature \textbf{462} (2009) 331-334.

\bibitem{Bolmont:2010np}
  J.~Bolmont and A.~Jacholkowska,
  Adv.\ Space Res.\  {\bf 47} (2011) 380.

\bibitem{Vasileiou}
V.~Vasileiou, A.~Jacholkowska, F.~Piron, J.~Bolmont, C.~Couturier, J.~Granot, F.~W.~Stecker, J.~Cohen-Tanugi and F.~Longo,
Phys. Rev. D \textbf{87} (2013) no.12, 122001.

\bibitem{Zhang2014}
S.~Zhang and B.~Q.~Ma,
Astropart. Phys. \textbf{61} (2015) 108-112.

\bibitem{xuma2016_2}
H.~Xu and B.~Q.~Ma,
Phys. Lett. B \textbf{760} (2016) 602-604.

\bibitem{IceCubeNature}
G.~Amelino-Camelia, G.~D'Amico, G.~Rosati and N.~Loret,
Nat.\ Astron.\ \textbf{1} (2017) 0139.

\bibitem{GACsmolinStaro}
G.~Amelino-Camelia, L.~Smolin and A.~Starodubtsev,
Class. Quant. Grav. \textbf{21} (2004), 3095-3110.

\bibitem{aschieri09}
P.~Aschieri, A.~Borowiec and A.~Pacho\l{},
[arXiv:2009.01051 [gr-qc]].

\bibitem{bianchiRovelli}
E.~Bianchi and C.~Rovelli,
Phys. Rev. D \textbf{84} (2011), 027502.

\bibitem{DSRFRW}
G.~Rosati, G.~Amelino-Camelia, A.~Marciano and M.~Matassa,
Phys.\ Rev.\ D \textbf{92} (2015) no.12, 124042.

\bibitem{planckParam}
N.~Aghanim \textit{et al.} [Planck],
Astron. Astrophys. \textbf{641} (2020), A6.

\bibitem{shaoLag}
L.~Shao \text{et al.},
Astrophys. J. \textbf{844} (2017) no.2, 126.


\end{thebibliography}
\end{document}